\documentclass[preprint,aps,amsmath]{revtex4}
\usepackage{graphicx}
\usepackage{amssymb}
\usepackage{bm}

\newcommand\bnabla{\boldsymbol \nabla}
\newcommand\pll{ \parallel }

\begin{document}

\title{Inertial blob-hole symmetry breaking\\in magnetised plasma filaments}

\author{Alexander Kendl}
\affiliation{Institute for Ion Physics and Applied Physics, University of
  Innsbruck, Association Euratom-\"OAW, Technikerstr. 25, 6020 Innsbruck,
  Austria\vspace{1cm}} 

\begin{abstract}
\vspace{1cm}
Symmetry breaking between the propagation velocities of magnetised plasma
filaments with large positive (blob) and negative (hole) amplitudes, as
implied by a dimensional analysis scaling, is studied with global (``full-n'')
non-Boussinesq gyrofluid computations, which include finite inertia effects
through nonlinear polarisation. 
Interchange blobs on a flat density background have higher inertia and
propagate more slowly than holes. In the presence of a large enough density 
gradient, the effect is reversed: blobs accelerate down the gradient and holes
are slowed in their propagation up the gradient. Drift wave blobs spread their
initial vorticity rapidly into a fully developed turbulent state, whereas
primary holes can remain coherent for many eddy turnover times.
The results bear implications for plasma edge zonal flow evolution and tokamak
scrape-off-layer transport.
\vspace{7cm}\\
{\sl This is a preprint version of a manuscript submitted to ``Plasma Physics and
  Controlled Fusion''.}
\end{abstract} 

\maketitle

\section{Introduction}

Localised pressure perturbations in magnetised plasmas are commonly refered to
as blobs for positive amplitudes $+\delta p$ in relation to the
background pressure $p$, and as holes for negative amplitudes $-\delta p$.
Pressure perturbations extend along the magnetic field lines into filaments,
whereas the motion perpendicular to the magnetic field ${\bf B}$ is
principally determined by drifts. 
In presence of a magnetic field inhomogeneity the curvature and gradient-B
drifts induce a dipolar potential structure, and the resulting electric
field's $E \times B$ drift drives the perturbations. 
Blobs travel down (and vice versa holes up) a mean gradient \cite{diamond95}.  

\smallskip

This propagation of blob filaments constitutes the major cross-field transport
mechanism for plasma density and heat in the scrape-off layer (SOL) of tokamak
fusion experiments, and is observed in many other magnetised laboratory or
natural plasmas \cite{krasheninnikov01,dippolito11,manz15}.
SOL interchange turbulence is considered to be composed of an ensemble of
nonlinearly interacting blobs and holes \cite{sarazin98}. 
Both blobs and holes appear to be primarily generated in the vicinity of the
separatrix, and can be experimentally identified by the sign of the skewness
in the probability distribution function of fluctuation amplitudes
\cite{boedo03,nold10}.  
The underlying interchange instability and the shape evolution of a plasma
blob into plumes share many features with the bouyancy driven Rayleigh-Taylor
instability in neutral fluids \cite{garcia06}. 

\smallskip

A common assumption presumes the propagation of blobs and holes to be
symmetric under simultaneous reversal of (C) the sign of the amplitude
(blobs or holes respectively) and of (P) the direction specified by
the gradient of the magnetic field. This combined CP symmetry is indeed well
fulfilled for blobs and holes with small amplitudes $\pm \delta p \ll p$
relative to the background plasma. 
Turbulence driven pressure fluctuations in the edge and SOL of magnetised
fusion plasmas however can have amplitudes in the order of unity 
\cite{endler95,zweben02,garcia04,garcia07,nold10}, 
and even above, as the propagation of edge-localised mode (ELM) 
filaments in the SOL is similar to large blobs \cite{kendl10}. 

\smallskip

The propagation and fragmentation of pressure perturbations like blobs is for
large amplitudes influenced by inertial (``global'') effects mediated mainly
through nonlinear polarisation \cite{wiesenberger14}. 
Most models and numerical codes for plasma blob
propagation and edge turbulence so far have however been making use of the
delta-n or Boussinesq approximation, which assumes small fluctuation amplitudes.
The inertial and nonlinear polarisation effects on drift wave turbulence and
blob propagation significantly modify the picture of edge and SOL
fluctuations. Recent results for global blob propagation obtained with
non-Boussinesq codes and models \cite{yu06,angus14,wiesenberger14,kube1112,manz13} 
demonstrate the relevance of full-n modelling in the edge  for more realistic
SOL blob transport scalings:  large inertial blobs are slowed on flat
background profiles, but accelerate strongly down pressure gradients.  

\smallskip

In light of these results on large inertial blobs it appears not at all any more
evident that large amplitude blobs and holes should be CP symmetric. 
The following work numerically studies asymmetries between inertial
interchange blobs and holes (2-d), and between large amplitude drift vortices
(3-d) with initially different polarities. 
The computational implementation is based on an isothermal reduction of the
full-n gyrofluid model by Madsen \cite{madsen13}, and reduces to the delta-n
model (GEM3) by Scott \cite{scott03,scott05} in the limit of small 
fluctuation amplitudes, and to the full-n model by Wiesenberger
\cite{wiesenberger14} in the two-dimensional limit. 

\smallskip

\section{Model and numerical methods}

The full-n 3-d gyrofluid model by Madsen \cite{madsen13} consists of a set of
6-moment equations and of the field equations for the potentials, completed by
a first order finite Larmor radius closure. 
In the following, an isothermal plasma is assumed, where temperature
variations in space and time are neglected. A normalised energetically
consistent set of 3-d full-n isothermal gyrofluid equations for electrons and
ions (species $s \in e,i$) for the first two moments (corresponding to eqs.~22
and 23 in ref.~\cite{madsen13}), which are the gyrocenter densities $n_s$ and
parallel velocities $v_{s}$, is: 
\begin{eqnarray} 
\partial_t \hat n_s &=&  {1 \over B} \left[\hat n_s, \phi_s \right] 
- {B \over n_s} \nabla_\pll  \left( { n_s v_s \over B} \right)
 + \kappa(h_s) \label{eq:den0}\\ 
\partial_t \alpha_s &=& {\mu_s \over B} \left[v_s, \phi_s \right]  
- \nabla_\pll h_s - C {J_\pll  \over n_s} 
 + \mu_s \tau_s v_s \kappa (\hat n_s) + 2 \mu_s \tau_s \kappa (v_{s}) 
\label{eq:vel0} 
\end{eqnarray}
with $\alpha_s \equiv ( \beta_0  A_{\pll} + \mu_s  v_{s})$ and 
$h_s \equiv  ( \phi_s + \tau_s {\hat n_s} )$.
Triple nonlinear terms including the parallel velocity are here neglected.
The nonlinear polarisation equation
\begin{equation}
\sum_s \left[ Z_s e \; \Gamma_{1s} n_s + \bnabla \cdot \left( n_s {|\mu_s| \over
    B^2} \bnabla \right) \phi \right] = 0.
\label{eq:pol0}
\end{equation}
determines the electrostatic potential $\phi$.
The gyro-screened potential is given by $\phi_s = \Gamma_{1s} \phi -
(\mu_s/2B) (\nabla \phi)^2$.
Parallel velocities and current are coupled to the vector potential
$A_{\pll}$ via Ampere's law 
$\nabla_{\perp}^2 A_\pll   =  - J_{||} = -\sum_s n_s Z_s e v_{s}$.
The gyro-averaging operator in Pad\'e approximation is defined by
$\Gamma_{1s} = (1 + (1/2) b_s)^{-1}$  with $b_s = \tau_s \mu_s \nabla_{\perp}^2$. 
The mass ratio is given by $\mu_s = m_s /(Z_s m_i)$, and the (constant)
temperature ratio by $\tau_s = T_s / (Z_s T_e)$. 
For electrons, thus $\tau_e = -1$, and finite Larmor radius (FLR) effects are
neglected ($b_e \equiv 0$). The electron contribution to the polarisation in
eq.~(\ref{eq:pol0}) is also neglected, as $|\mu_e| \ll |\mu_i|$.
The gyrocenter densities $n_s$ are normalised to a constant reference density
$n_0$, so that the magnitude of the plasma density $n_s \leftarrow n_s/n_0$ 
is of order one. Eqs.~(\ref{eq:den0}, \ref{eq:vel0}) have been divided by the
specific variable densities $n_s$, and logarithmic densities 
$\hat n_s \equiv \ln n_s$ are introduced to ensure positivity, with both
${\hat n_s}$ and $n_s$ appearing in the equations. 

\smallskip

The spatial derivative operators are normalised as $\nabla \leftarrow \rho_s \nabla$
to the drift scale $\rho_s = (c/eB_0) \sqrt{m_i T_e}$, where  $m_i$ is the
mass of the main ion species, $T_e$ is a constant reference electron
temperature, and $B_0$ is a static reference background magnetic field strength.  
Parallel derivatives are further scaled as $\nabla_\pll \leftarrow
(L_\pll/L_{\perp}) \nabla_\pll$ with the connection length $L_\pll$, which for
toroidal geometry is given by $L_\pll = 2 \pi q R$ with inverse rotational
transform $q$ and major torus radius $R$. 
The drift parameter $\delta = \rho_s / L_\perp$ is used to set the
perpendicular length scale $L_{\perp}$.  
For blob simulations often $L_{\perp} = \rho_s$ is used (so that $\delta=1$),
and for gradient driven turbulence usually $L_{\perp} = L_n$ is set as the
density gradient length scale $L_n$. In order to apply the same normalisation
length for all presented simulations (including those on drift wave vortices),
a normalisation to a typical edge gradient length is chosen with $\delta=0.01$.

\smallskip

The time scale is normalised as $\partial_t
\leftarrow (\rho_s / c_s) \partial_t$, and parallel velocities $v_{\pll s}
\leftarrow v_{\pll s}/c_s$ are normalised by the sound speed  $c_s =
\sqrt{T_e/m_i}$. 
Further, $\phi \leftarrow (e \phi/T_e)$, $B \leftarrow B/B_0$, 
$J_\pll \leftarrow J_\pll /(e n_0 c_s)$,
and $A_\pll \leftarrow (A_\pll/\beta_0 B_0 \rho_0)(L_\perp/qR)$ for a reference
electron beta given by $\beta_0 = 4\pi n_{0} T_{e0} / B_0^2$.
The collisionality parameter is given by $C = (L_\perp/ c_s \rho_0 B_0) \eta$
with $\eta = 0.51 (m_e \nu_e) / (n_0 e^2)$.
The main plasma parameters are  ${\hat  \mu}_s = \mu_s {\hat \epsilon}$, 
$\hat \beta = (n_{0} T_e / B_0^2) \hat \epsilon$, and 
$\hat C = 0.51 (m_e \nu_e L_{\perp} / c_{s0}) \hat \epsilon$ 
with ${\hat \epsilon} = (qR/L_{\perp})^2$. In the following only electrostatic
blobs and vortices with $\hat \beta=0$ are discussed (while electromagnetic
effects are of more relevance for fully developed turbulence).

\smallskip

The 2-d advection terms are expressed through Poisson brackets
$[f,g] = (\partial_x f)(\partial_y g) - (\partial_y f)(\partial_x g)$ 
for locally perpendicular coordinates $x$ and $y$.   
Normal and geodesic components of the magnetic curvature enter the
compressional effect due to field inhomogeneity by  
$\hat \kappa = \hat \kappa_y  \partial_y + \hat \kappa_x  \partial_x$
where the curvature components in toroidal geometry
are functions of the poloidal angle $\theta$ mapped onto the parallel
coordinate $z$.  
For a circular torus $\hat \kappa_y \equiv \kappa_0 \cos(z)$ and $\hat
\kappa_x \equiv \kappa_0 \sin(z)$ when $z=0$ is defined at the outboard midplane.  
An Arakawa-Karniadakis numerical scheme
\cite{arakawa66,karniadakis91,naulin03} is used for the computation  
of eqs.~(\ref{eq:den0}) and (\ref{eq:vel0}). The generalised Poisson type
equation (\ref{eq:pol0}) is solved by a Chebyshev accelerated 4th order
red-black SOR scheme \cite{leveque,nagel14,humphries}.
For numerical stability, a small perpendicular hyper-viscosity term $s_\nu = - \nu_4
\nabla_{\perp}^4 \hat n_s$ is added on the right hand side of
eq.~(\ref{eq:den0}), and in 3-d computations parallel viscous terms
$\nu_\pll \partial_z^2 \hat n_s$ and $\nu_\pll \partial_z^2 v_s$ are added to 
eqs.~(\ref{eq:den0}) and (\ref{eq:vel0}), respectively.

\smallskip

Boundary conditions in $y$ direction are periodic for 2-d simulations, and
quasi-periodic (shear-shifted flux tube) for 3-d simulations.
The total density is allowed to evolve freely, although for the present short
blob propagation times the initial background profiles do not evolve
visibly. To avoid degradation and flows at the radial boundaries, fixed mixed (von
Neumann / Dirichlet) vorticity free ($n_e = \Gamma_1 n_i$) boundary
conditions are applied in $x$. For longer turbulence simulations with free
profile evolution, sources and sinks would rather have to be specified at
the radial boundaries. 

\smallskip

The delta-n isothermal electromagnetic gyrofluid model \cite{scott05,kendl14} 
is regained by splitting $n_s = n_{s0} + \tilde n_s$ into a static constant
background density $n_{s0}$ and the perturbed density $\tilde n_s$. 
When $\tilde n_s / n_{s0} \ll 1$, the right hand sides of eqs.~(\ref{eq:den0}) and
(\ref{eq:vel0}) can be linearised by approximating 
$n_s \approx n_{s0}$ so that $\hat n_s \approx \hat n_{s0}
+ ({\tilde n}_s / n_{s0})$, and neglecting all nonlinear terms except the
Poisson bracket:
\begin{eqnarray} 
\partial_t \tilde n_s &=&  {1 \over B} [ \tilde n_s, \tilde \phi ]
- B \nabla_\pll  \left( { \tilde v_{s} / B} \right) 
+ \hat \kappa(h_s) \label{eq:denloc}\\ 
\partial_t \alpha_s  &=& {\mu_s \over B} [ \tilde v_s, \tilde \phi ]  
- \nabla_\pll h_s + 2 \mu_s \tau_s {\hat \kappa} (\tilde v_{s}) - C J_\pll \label{eq:velloc}
\end{eqnarray}

\smallskip

The consistent delta-n polarisation equation in the high-$k$ limit is
\begin{equation}
\sum_s  a_s [ \Gamma_{1s} \tilde n_s + (1/\tau_s) (\Gamma_{0s} -1)
  \tilde \phi ]  =  0
\end{equation}
with  $\Gamma_{0s} = (1 + b_s)^{-1}$.
Linearisation of the low-$k$ eq.~(\ref{eq:pol0}) actually does not include
the gyro-screening on the potential and results in 
$\sum_s  a_s \Gamma_{1s} \tilde n_s  = \nabla_{\perp}^2 \tilde \phi$.  
The velocities and current are again coupled to the parallel component of the
fluctuating vector potential by Ampere's equation
$\nabla_{\perp}^2 \tilde A_{||}   =  \tilde J_{||}  =  \sum_s a_s \tilde v_{s}$.
The parameter $a_s = Z_s n_{s0}/ n_{e0}$ describes the ratio of species
reference densities $n_{s0}$ to $n_{e0}$.

\section{Large inertial 2-d interchange blobs and holes}

In the following, large amplitude blob and hole propagation is compared for
the full-n and delta-n models. To separate 2-d interchange and 3-d drift wave
effects, at first the computations are restricted to 2-d by neglecting the
parallel velocity and parallel derivatives. In this limit the equations
correspond to the 2-d full-n model by Wiesenberger \cite{wiesenberger14}. 

\smallskip

Blobs and holes are initialised as Gaussian density perturbations with width
$r = 10 \rho_s$ and amplitude $\Delta n = \pm 0.75$ for $n_{b} = 1$. 
In the full-n model $n_{b}$ corresponds to the actual background plasma
density, whereas in the delta-n model this can be regarded as a dummy
parameter on which the solution does not depend, as the model already implies
a large underlying background $n_{0} \gg \Delta n$.
Dimensional analysis roughly estimates the delta-n and full-n blob propagation
speed scalings \cite{kube1112,wiesenberger14} as 
\begin{eqnarray}
{V_{delta} \over c_s} &\sim& \sqrt{\Delta   n}, \hspace{0.5cm}\mbox{and}\\
{V_{full} \over c_s} &\sim& \sqrt{\Delta n \over (n_b+\Delta n)}.
\end{eqnarray}

On this basis inertial blobs could be expected to propagate more slowly than
holes (with reverse direction) in the full-n model.  

\begin{figure} 
\hspace{-0.5cm}
\includegraphics[width=10.5cm]{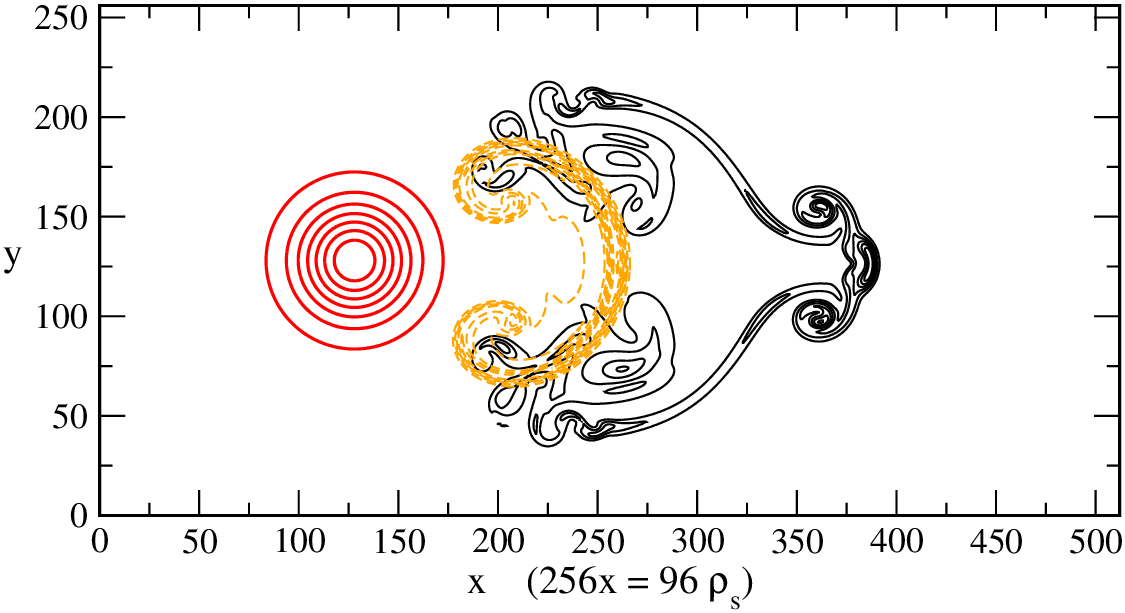}\\
\includegraphics[width=11.5cm]{red-fig-1bc.epsi}
\vspace{5mm} 
\caption{\sl Top (delta-n): Large amplitude blob/hole evolution in the delta-n
  model at t=0 
  (red bold), t=12.5 (orange dashed) and t=25 (black thin line). The contours
  of delta-n blob and hole coincide for simultaneous reversal of amplitude and
  x-direction. Bottom (full-n):  different states of inertial blob (left) and
  hole (right: x-direction  mirrored) at t=25.\vspace{10mm}}  
\label{f:fig-flat}
\end{figure}

\smallskip

Simulation parameters here are $\kappa = 0.05$, $\delta=0.01$ and $\tau_i=0$. 
The computational grid is $n_x \times n_y = 512 \times 256$ with resolution
$(192 \times 96) 
\rho_s$. Fig.~\ref{f:fig-flat} (top) shows the symmetric evolution of blobs
and holes in the delta-n model (top) for times $t=0$, $t=12.5$ and $t=25$. For simultaneous
reversal of grad-B direction and sign of the amplitude, delta-n blob and hole coincide:
the density contours are identical for blobs and (reversed) holes.
The bottom figures show the different states at $t=25$ for an inertial
full-n blob (left) and full-n hole (right). The inertial blob has a more coherent
head and, propagates slower than for the delta-n case, whereas the inertial
hole fragments more strongly and propagates faster, as predicted by the inertial scaling.

\smallskip

\begin{figure} 
\includegraphics[height=6.5cm]{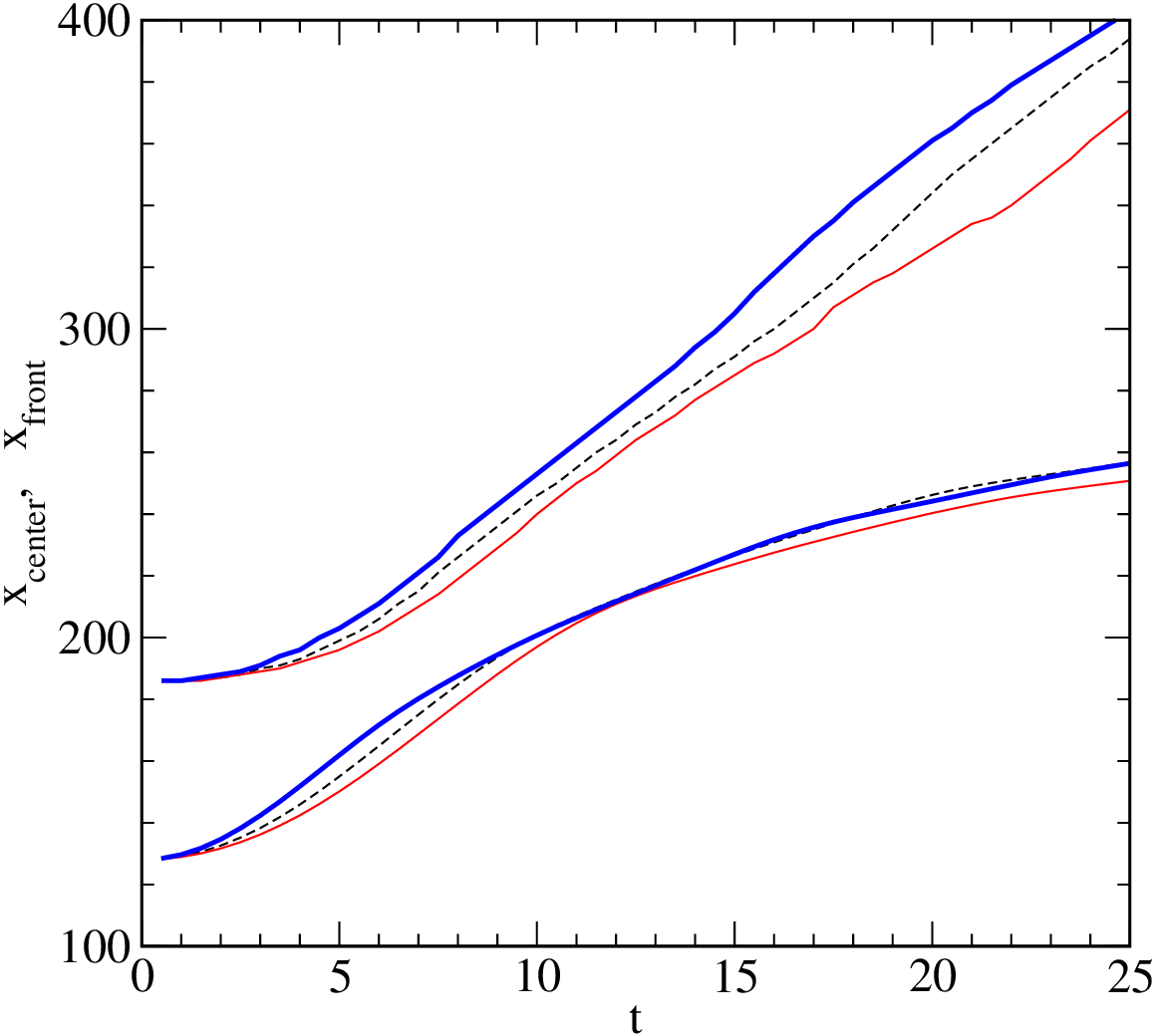}
\includegraphics[height=6.5cm]{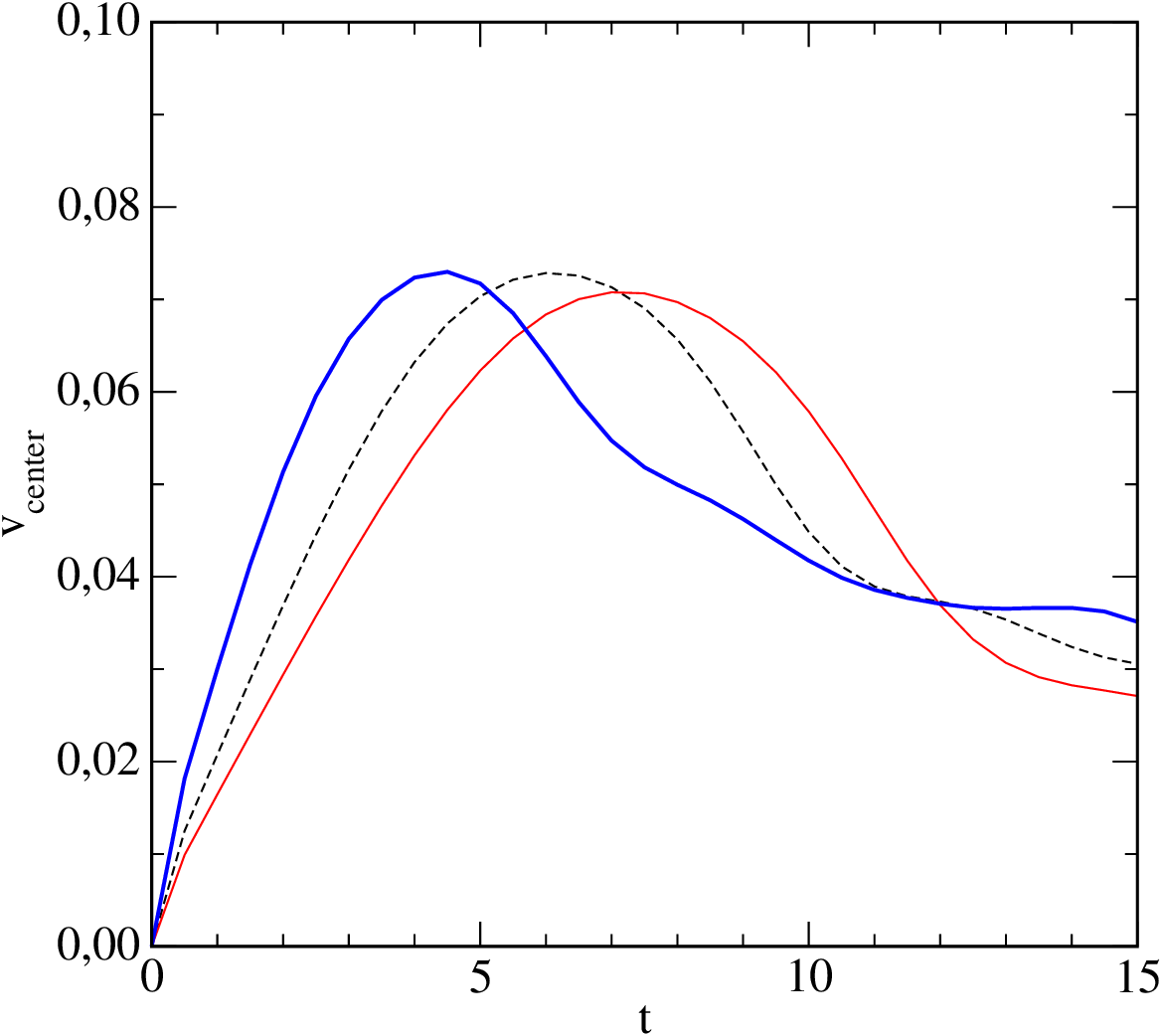}
\caption{\sl Left: time evolution of the center of mass (bottom lines) and
  propagation fronts (upper lines) for a delta-n blob (dashed black),
  inertial blob (thin red) and inertial hole (bold blue). Right: center of
  mass speeds. (Time in units $L_{\perp}/c_s$)}  
\label{f:fig-fronts}
\end{figure}

Fig.~\ref{f:fig-fronts} (left) shows the corresponding time evolution of the
x-coordinates of the center of mass (bottom lines) and the propagation fronts
(upper lines) for the delta-n case (black dashed lines), the inertial blob
(thin red lines) and inertial hole (bold blue lines, mirrored in x direction). 

\smallskip

The radial center of mass position is determined by
\begin{equation}
x_{center} = \left| { \sum_{i,j} x_i [ n_e(x_i,y_j) - n_b(x_i) ]  \over 
\sum_{i,j} [ n_e(x_i,y_j) - n_b(x_i) ] } \right|
\end{equation}
and the center of mass velocity by $v_{center} = \Delta x_{center} / \Delta t$.
The blob front position is here simply determined as the furthest outward $x$
position where the density deviates more than 10~\% from the initial background profile. 

The acceleration occurs mostly in the initial quasi-linear phase (compare
center-of-mass velocity plots in right figure), while at later times the
center-of-mass velocities drop and the front velocities saturate nearly equally
for blobs and holes.

\smallskip

The maximum center of mass velocity in general depends on the initial blob amplitude
and width, which are here kept fixed. For given width and amplitude, the
maximum velocities are found to be similar, with a slightly reduced maximum
velocity for the inertial (full-n) blob compared to the delta-n blob, and a
slightly increased velocity for the inertial hole. This observation is
consistent with results for large amplitude blobs presented in
ref.~\cite{wiesenberger14}. 

\smallskip

So far a constant background density has been assumed. Now a linearly decreasing
background density profile $n_b(x) = 2(1-x/x_{max})$ is considered. 
Blobs thus propagate into regions of lower background density, and holes into
higher density.  
In a delta-n model the blob/hole velocity would be unchanged. 
Inertial blobs with 
\begin{equation}
{V_{full} \over c_s} \sim \sqrt{\Delta n \over (n_b(x)+\Delta n)}
\end{equation}
however can be
expected to accelerate, and holes to be slowed down. For large enough
gradients the inertial effects on blob/hole velocities found for flat profiles
can even be reversed.

\smallskip

\begin{figure} 
\includegraphics[height=4.2cm,width=14.2cm,angle=0]{red-fig-3-a_grad-bn.epsi}\\
\hspace{-17mm}\includegraphics[width=13.0cm]{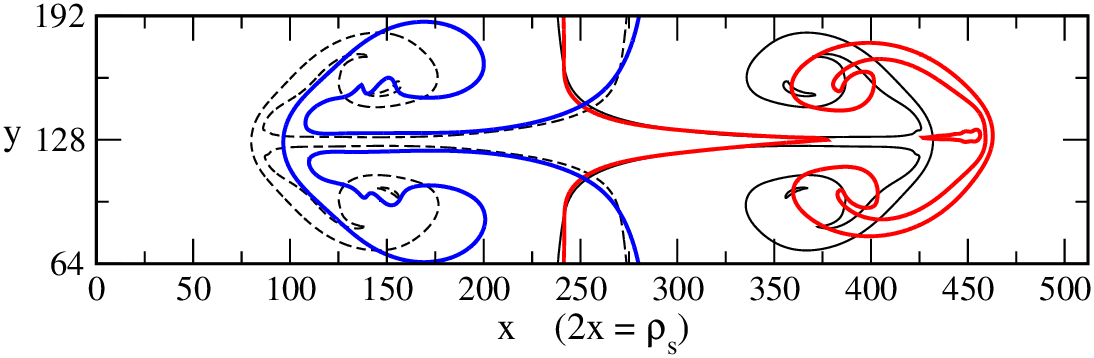}\\
\caption{\sl Top: delta-n blob evolved until $t=15$ on a density gradient.
Bottom: $n=1$ density contour plots of the delta-n blob (thin black line), 
anti-symmetric delta-n hole (dashed black line), accelerated full-n inertial
blob (bold red line) and decelerated full-n hole (bold blue line).}
\label{f:fig-grad}
\end{figure}

This is demonstrated in computations with $\Delta n = \pm
0.85$ and resolution $(128 \times 256) \rho_s$, with the initial blob/hole
located in the middle of the domain, for otherwise identical parameters, in
Fig.~\ref{f:fig-grad}. 
The top figure shows a delta-n blob propagating down a density gradient at $t=15$.
In the bottom density contour plots (at $n=1$) of the same delta-n blob (thin
black line) and its anti-symmetric delta-n hole (dashed black line) are shown,
together with a full-n inertial blob (bold red line) that has accelerated
further down the gradient (i.e. to the right side) and a full-n hole (bold
blue line) which is slowed during propagation into denser regions. 

\smallskip

To sum up these first results, in a delta-n model 2-d interchange blobs and
holes evolve identically and regardless of a background density gradient. 
In the inertial full-n model, on a constant background the blobs move more
slowly and coherently, and the holes faster and more fragmented. 

\smallskip

On the other hand, on a background density gradient the inertial full-n holes,
which move up the gradient, decelerate, whereas blobs accelerate down the
gradient. The relative evolution and propagation of negative and positive
perturbations thus strongly depends on the background gradient. 
Gradient steepening around the separatrix (where blobs and holes are most
likely born) accordingly will lead to very different transport behaviour for
full-n (or non-Boussinesq) models compared to results obtained in delta-n models. 

\smallskip

\begin{figure} 
\includegraphics[width=12.0cm,angle=0]{red-fig-4ab.epsi}\\
\includegraphics[height=5.0cm]{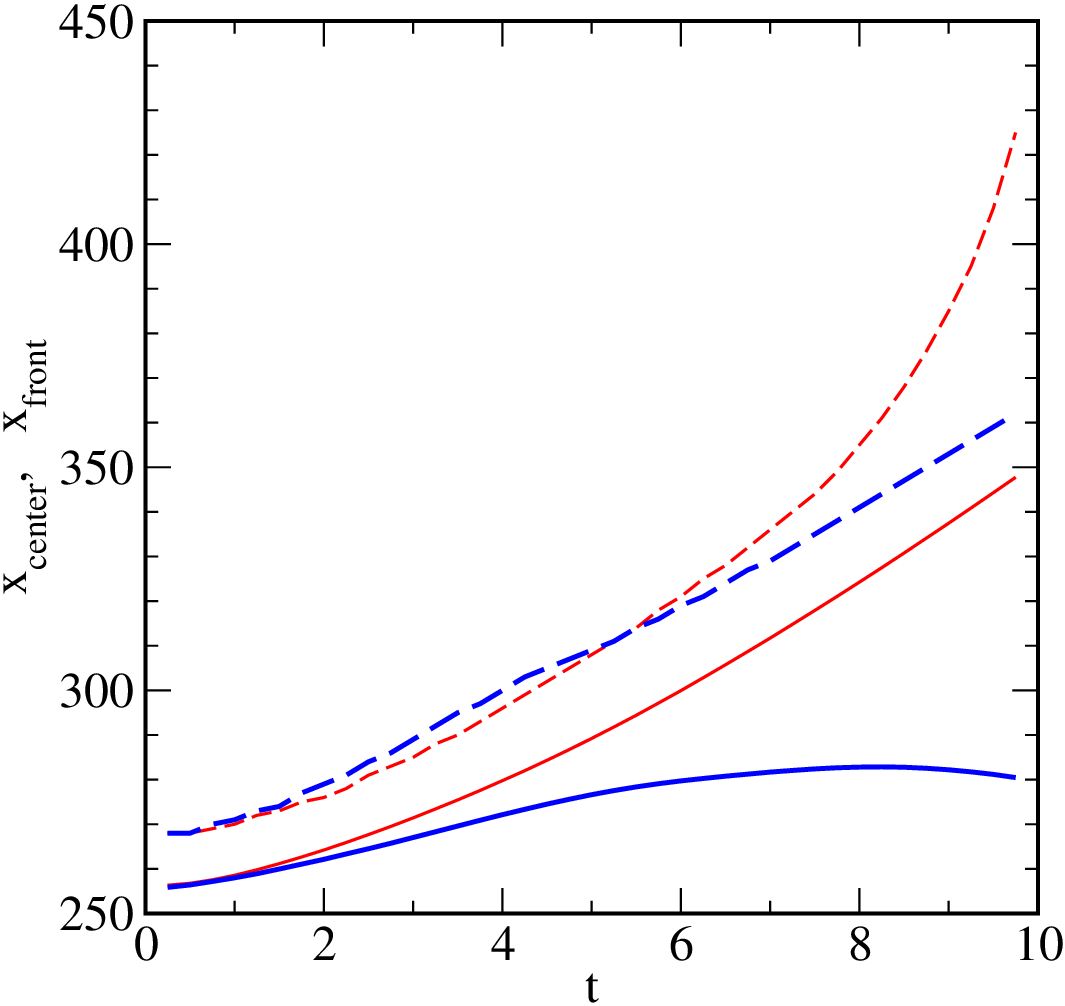}
\includegraphics[height=5.0cm]{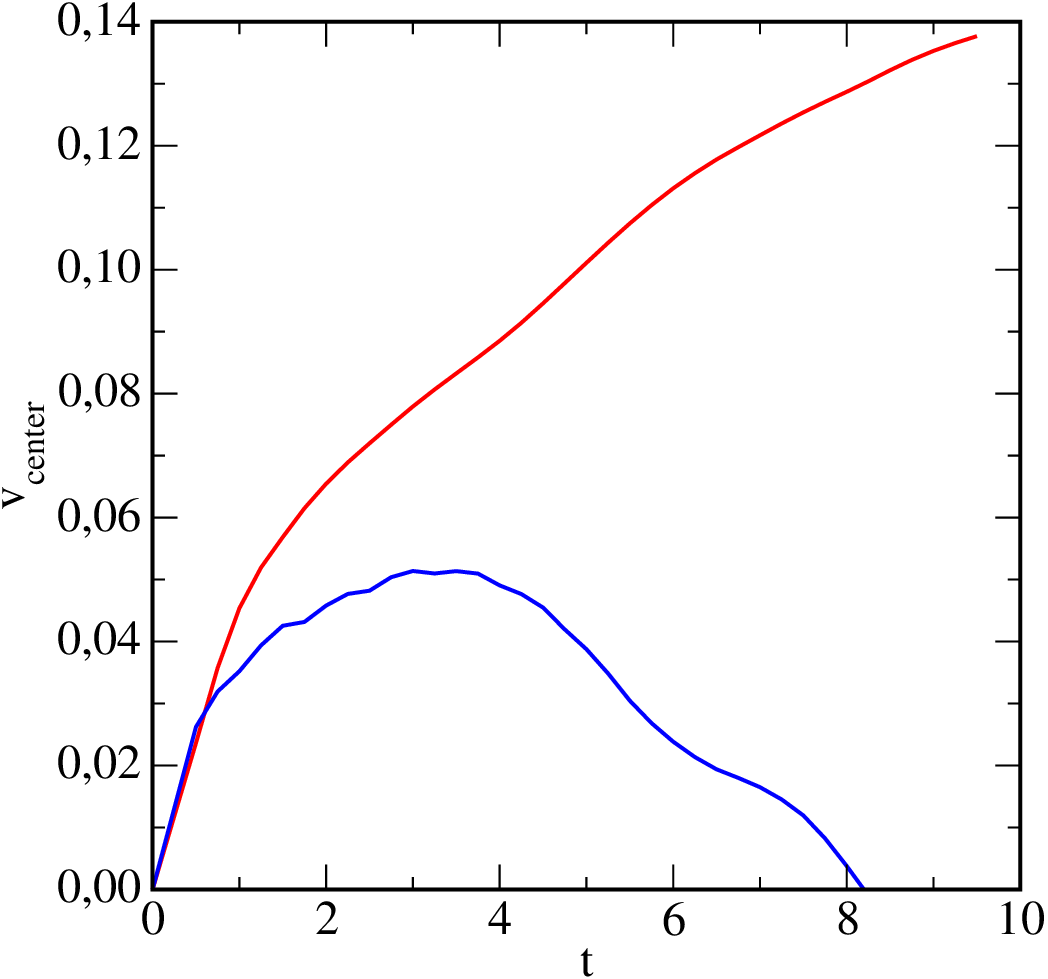}
\vspace{0.5cm}
\caption{\sl Top: Inertial large warm ion ($\tau_i=2$) hole (left) and blob
  (right) on a density gradient: the more coherent head shows complicated
  poloidal-radial  propagation. 
Bottom left: time evolution of the center of mass (bottom solid lines) and
  propagation fronts (upper dashed lines) for a inertial blob (red) and hole
  (blue, with x-direction inverted). Bottom right: corresponding center of
  mass veleocities for the blob (red upper curve) and hole (blue lower
  curve).}
\vspace{0.5cm}
\label{f:fig-tau2}
\end{figure}

\vspace{3.0cm}
Next, the inertial evolution for warm ions with $\tau_i = 2$ is considered,
which is a typical value for the tokamak SOL. Warm ions primarily enhance the
blob 
propagation speed \cite{manz13} by contributing to the interchange drive,
break the (approximate) up-down symmetry through FLR effects on polarisation,
and remain more coherent \cite{madsen11,wiesenberger14}. Although the
radial and poloidal propagation is more complicated (the hole head
e.g. changes direction twice within the computation time), the major
conclusions remain (Fig.~\ref{f:fig-tau2}): the inertial large blob
center-of-mass velocity is larger than for the hole. The front velocity of the
hole is initially higher than for the blob, but is reduced after sufficient
propagation into the denser region.
%

\bigskip

\section{Large inertial 3-d blobs, holes and drift vortices}

3-d field-aligned computations of blobs and holes include different parallel
electron and ion dynamics, which introduces charging and polarisation of the
pressure perturbation, resulting in Boltzmann spinning of the blob \cite{angus12}.  

\smallskip

The charging of a blob, which is initially localised in parallel direction, is
a consequence of the higher parallel mobility of the electrons. 
In eq.~(\ref{eq:vel0}) the acceleration $\partial_t v_s \sim 1/\mu_s$ is inversely
proportional to the species mass, so that the resulting parallel current 
$J_{||} \approx n_e e v_e$ is mostly carried by the electrons.
In the absence of collisions ($C=0$), electrons tend, according to the parallel
component of eq.~(\ref{eq:vel0}), towards a Boltzmann response with
$\nabla_{||}h=0$, so that the electrostatic potential $\phi \sim \hat n_e$
spatially aligns with the blob. The resulting $E \times B$ drift leads to a
perpendicular spinning vortex around the blob. 
For finite collisionality $C>0$ the relative importance of spinning is
controlled by the balance between the divergence of the parallel current and
the divergence of the diamagnetic and polarisation currents under
quasi-neutrality. The Boltzmann charging then is reduced, and depending on $C$
the radial interchange drive competes with poloidal drift wave motion.

\smallskip

The present computations show that the spin-up of blob rotation by 3-d drift
wave dynamics is strongly dependent on the collisionality parameter: for
typical edge pedestal values in the closed-flux-surface region of $\hat C =
3.5$ the Boltzmann charging is 
dominant (cf. \cite{angus12,angus14}), but for an order of magnitude larger
values ($\hat C \sim 20-50$), as more appropriate for mid-SOL plasmas, the
interchange drive and the typical 2-d like blob plume structure actually
prevail. In the presence of a density gradient, drift wave type propagation in
the electron diamagnetic direction and instability add to the dynamics. 

\smallskip

First, the flat background profile case with $\tau_i=0$ (as in
Fig.~\ref{f:fig-flat}) is re-considered by extending the otherwise same
computation to $n_z=16$ planes in the field-aligned direction, including
consistent poloidal (parallel) variation of the background magnetic field
gradient $\kappa(z)$.  
The initial background density here is set constant in the parallel direction,
and the initial electrostatic potential and parallel velocities are zero.
The blobs are initially localised in the middle $z$ plane.

\smallskip

\begin{figure} 
\hspace{-0.5cm}
\includegraphics[width=12.0cm,angle=0]{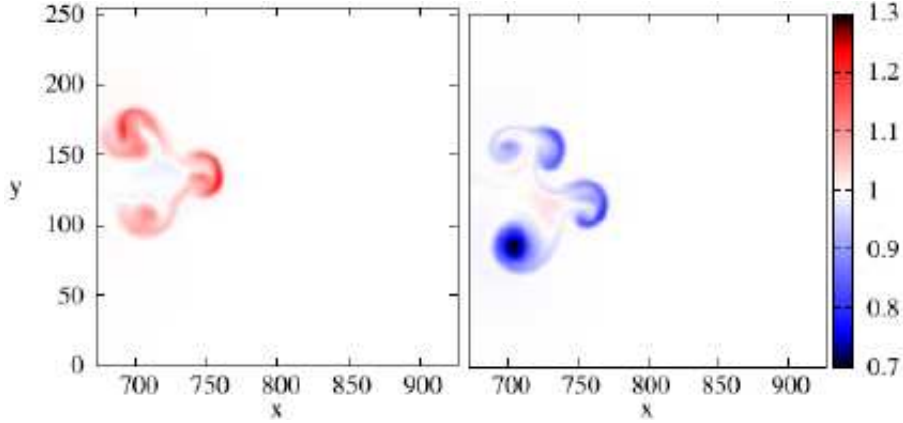}\\
\caption{\sl 3-d computations at $t=25$ of inertial blob (left) and hole (right, x-axis
  inverted) for $\hat C=7.5$ and $\tau_i=0$ (and otherwise same parameters as
  for Fig.~\ref{f:fig-flat}) show strong Boltzmann spinning (which is
  significantly reduced for higher collisionalities).}   
\label{f:fig-3d}
\end{figure}

For $\hat C = 7.5$ the Boltzmann spinning effect indeed is well pronounced, as
shown in Fig.~\ref{f:fig-3d}: the radial propagation is reduced compared with
the 2-d (or a more strongly resistive) case.  
Holes charge up negatively and blobs positively, and obtain opposite Boltzmann
spins. As the head is accordingly rotationally advected, the upper arm of the blob and the
lower arm of the hole get more pronounced, respectively. It is also 
observed that the spinning hole shows stronger coherence than the blob.

\smallskip

Finally, the 3-d evolution of large amplitude ($\Delta n = \pm 0.85$) drift
wave blobs and holes in a sheared slab geometry with $\hat s =1$,  $\kappa=0$, $\hat
C = 3.5$ and $\hat \epsilon = 18000$ on a background edge density gradient
$n_b(x)=1.5-x/x_{max}$ is studied.  
The simulation domain is $(n_x \times n_y \times n_z) = (96 \times 256 \times 16)\rho_s$.
Drift wave blobs show a rapid transition into fully developed turbulence. 
Here only the initial stage is considered. 

\smallskip

For small amplitudes (or in a delta-n model) the development of nonlinear
drift vortices is exactly CP-symmetric for initial blobs compared to holes
(up to computing precision): the spatio-temporal contours are identical for
reversal of the density gradient direction (P), while density fluctuation, potential
and vorticity amplitudes are also reversed (C). 

\smallskip

\begin{figure} 
\includegraphics[width=12.0cm,angle=0]{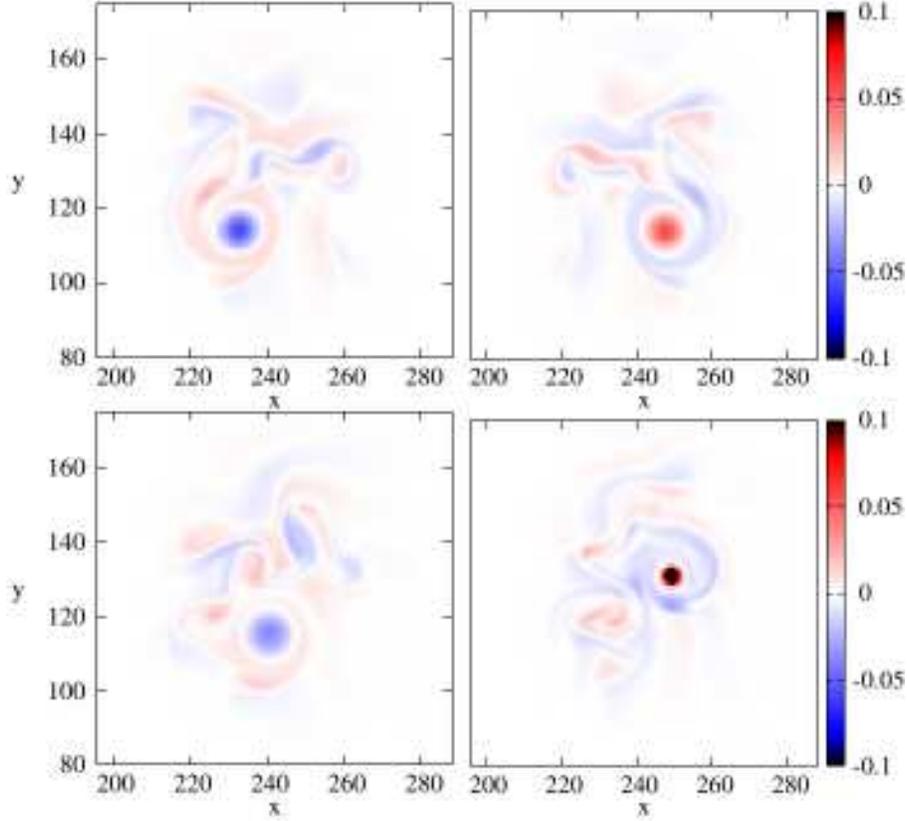}\\
\vspace{0.5cm}
\caption{\sl Vorticity $\Omega$ of 3-d drift vortices evolved at $t=50$ from large
  positive and negative density perturbations: in the delta-n model (top row)
  the blob (left) and hole   (right) drift vortices are perfectly
  anti-symmetric. In the full-n model (bottom row) the negative initial
  vorticit of the blob has spread and its amplitude decreased. The positive
  vorticity of the initial hole is spatially compressed at increased
  amplitude. (Only part of the $(x,y)$ domain at $z=8$ is shown.)\vspace{0.5cm}
}   
\label{f:fig-dw}
\end{figure}

Fig.~\ref{f:fig-dw} on top shows the vorticity $\Omega = \nabla_{\perp}^2
\phi$ of blob (left) and hole (right) delta-n drift vortices at $t=50$.
Large drift wave blobs and holes however show different evolution in the
consistent full-n model: the primary blob vortex (Fig.~\ref{f:fig-dw} bottom
left) has spread and its amplitude is decreased compared to the delta-n case,
whereas the hole vortex (bottom right) is compressed radially with a strongly
increased vorticity amplitude. The hole actually can be observed as a coherent
tripolar vortex for quite some time (multiple eddy turnover times) during the
development into a fully turbulent state of the secondary drift wave structures. 

\smallskip

As drift wave turbulence in the outer closed-flux-surface edge pedestal region
near the separatrix can acquire fluctuations amplitudes in the same order of
magnitude as the background, these results also show the relevance of full-n
models for edge turbulence (and probably for the understanding of edge
transport barriers) in addition to the relevance for modelling of SOL blobs
and interchange turbulence.

\vspace{3cm}
\section{Conclusions}

To summarise, symmetry breaking between the evolution of magnetised plasma
filaments with large positive (blob) and negative (hole) amplitudes has been
found. Interchange blobs on flat density background have higher inertia than
holes and propagate more slowly. In the presence of a large enough density
gradient, blobs accelerate down the gradient and holes are slowed in their
propagation up the gradient. Gradient steepening at the blob/hole birth region
(supposedly near the separatrix) can thus lead to enhanced blob velocities and
transport into the outer SOL. This mechanism would be consistent with
observations at various tokamaks on an effect of core density increase on
flattening of the outer SOL profiles \cite{labombard01,carralero14}.
Another implication is that in the presence of a strong background gradient
the inward impurity convection across the separatrix by holes can be
reduced, and alignment of (trace and non-trace) impurities in
vortices \cite{kendl12} can be expected to be significantly modified.

\smallskip

Full-n effects on large amplitude edge turbulence vortices, as they were
demonstrated in this work, can lead to profound consequences. 
For example, large inward propagating holes can remain coherent on a turbulent
background for significant times. It would be possible for such holes to be
trapped on resonant surfaces (where they would not be filled up rapidly by
parallel connection) and rotate for longer times with the background
plasma. This could explain phenomena like palm tree modes
\cite{koslowski05}. 

\smallskip

Most of all, strong effects on the generation and structure of zonal flows
(and, supposedly, mean flows) can be expected.  
As ion temperature fluctuations in the SOL also can achieve large amplitudes
\cite{kocan12}, both SOL and edge turbulence have to be studied with more complete
source-driven full-n gyrofluid models including temperature and heat 
transport equations (or full-f gyrokinetic equations \cite{scott10}), and with consistent
coupling to the SOL including appropriate sheath boundary conditions. 
Such models are presently under development. 
The presented results clearly show the necessity for full-n, non-Boussinesq
turbulence and blob transport models for the tokamak edge/SOL region. 

\vspace{3cm}
\section*{Acknowledgements}

The author thanks J. Madsen (DTU), M. Held and M. Wiesenberger (U Innsbruck) for
useful discussions on full-n gyrofluid models. 
This work was partly supported by the Austrian Science Fund (FWF) Y398.
This work has been carried out within the framework of the 
EUROfusion Consortium and has received funding from the 
Euratom research and training programme 2014-2018 under grant
agreement No 633053. The views and opinions expressed herein
do not necessarily reflect those of the European Commission.”

\end{document}